\documentclass[prd,amsmath,amssymb]{revtex4}

\usepackage{graphicx}
\usepackage{bm}

\normalbaselines \topmargin -0.25 truein \oddsidemargin 0.30
truein \evensidemargin 0.30 truein 

\begin{document}

\title{Extended Einstein-Dirac-aether theory: \\ Spinor modifications of the dynamic aether kinetic term}

\author{Alexander B. Balakin}
\email{Alexander.Balakin@kpfu.ru} \affiliation{Department of General
Relativity and Gravitation, Institute of Physics, Kazan Federal University, Kremlevskaya
str. 18, Kazan 420008, Russia}
\author{Anna O. Efremova}
\email{anna.efremova131@yandex.ru} \affiliation{Department of General
Relativity and Gravitation, Institute of Physics, Kazan Federal University, Kremlevskaya
str. 18, Kazan 420008, Russia}

\date{\today}

\begin{abstract}
In the framework of the Einstein-Dirac-aether theory the extended model is established, in which the backreaction of the spinor field on the unit vector field, describing the dynamic aether, is taken into account. For this purpose, we extended the Jacobson constitutive tensor, which enters the kinetic part of the Lagrangian of the unit vector (aetheric) field, by the terms containing spinor tensors and pseudotensors. The self-consistent system of master equations for the unit vector field, as well as, the spinor and gravitational fields is derived using the variation formalism in the framework of the effective field theory up to the second order in derivatives.
\end{abstract}

\maketitle

\section{Introduction}

The theory of dynamic aether formulated in \cite{J1} was developed later in many works (see, e.g., the status report \cite{J2}). One of the trends in the aether theory extensions deals with the interactions of the dynamic aether with the electromagnetic, pseudoscalar and spinor fields \cite{BL,B,BS,BE1,BE2}. In this context one can highlight the idea that the dynamic aether controls the evolution of mentioned fields (see, e.g., \cite{BS}). Now we are interested in studying the backreaction of the spinor field on the dynamic aether. In \cite{BE3} we discussed the variant of a backreaction based on the analysis of the potential of the axion field, which depends on the scalars constructed on spinor and unit vector (aether) fields. Now, in this short communication we extend the kinetic part of the unit vector (aether) field by including terms depending on the  spinor field; in this study we restrict ourselves by the terms up to the second order in derivatives in the context of effective field theory. Such approach extends the arsenal of instruments of this theory, since now in the process of reconstruction of the Jacobson constitutive tensor we can use not only the metric coefficients and velocity four-vector, but the spinor tensors and spinor pseudotensors in convolutions with the Levi-Civita pseudotensor.

Below we reconstruct the total ten-parameter Lagrangian of the model, and derive the self-consistent system of extended master equations for the unit vector, spinor and gravitational fields.

\section{The formalism}

\subsection{Action functional of the canonic Einstein-Dirac-aether theory}

The total action functional of the canonic Einstein-Dirac-aether theory is considered to be of the form
$$
S = \int d^4x\sqrt{{-}g}\left\{{-}\frac{1}{2\kappa}\left[R{+}2\Lambda {+} \lambda \left(g_{mn}U^m
U^n {-}1 \right) {+}K^{abmn} \nabla_a U_m  \nabla_b U_n  \right] {+} \right.
$$
\begin{equation}
\left. + \frac{i}{2}[\bar{\psi}\gamma^{k}D_{k}\psi {-}D_{k}\bar{\psi}\gamma^{k}] {-} m \bar{\psi}\psi \right\} \,.
\label{1}
\end{equation}
The elements of the canonic Lagrangian of the Einstein-aether model are well documented. The term $g$ is the determinant of the spacetime metric $g_{mn}$; $R$ is the Ricci scalar; $\Lambda$ is the cosmological constant; $\kappa {=} 8\pi G$ includes the Newtonian coupling constant $G$ ($c{=}1$).
The four-vector $U^j$ is associated with the aether velocity; the term $\lambda \left(g_{mn}U^m U^n {-}1 \right)$ designed to guarantee that the $U^j$ is normalized to one;  $\lambda$ is the Lagrange multiplier. The kinetic term $K^{abmn} \ \nabla_a U_m \ \nabla_b U_n $ is quadratic in the covariant derivative $\nabla_a U^m $ of the vector field $U^m$.
The spinor field with the mass $m$ is described by two terms $\psi$ and $\bar{\psi}$; $\gamma^k$ are the Dirac matrices; the covariant derivatives of the spinor fields
\begin{equation}
D_{k}\psi = \partial_{k}\psi-\Gamma_{k}\psi \,, \quad
D_{k}\bar{\psi}=\partial_{k}\bar{\psi}+\bar{\psi}\Gamma_{k} \,,
\label{D}
\end{equation}
are constructed using the Fock-Ivanenko connection matrices $\Gamma_{k}$.
The canonic constitutive tensor $K^{abmn}$ is constructed
using the metric tensor $g^{kj}$ and the aether velocity four-vector $U^k$ only:
\begin{equation}
K^{abmn} {=} C_1 g^{ab} g^{mn} {+} C_2 g^{am}g^{bn}
{+} C_3 g^{an}g^{bm} {+} C_4 U^{a} U^{b}g^{mn}.
\label{2}
\end{equation}
The parameters $C_1$, $C_2$, $C_3$ and $C_4$ are the Jacobson coupling constants.

\subsection{Extension of the Lagrangian of the Einstein-Dirac-aether theory}

In order to extend the kinetic part of the Lagrangian (\ref{1}), we replace the term $K^{abmn} \nabla_a U_m  \nabla_b U_n $ with
${\cal K}^{abmn} \nabla_a U_m  \nabla_b U_n $, where the extended constitutive tensor is
\begin{equation}
{\cal K}^{abmn} = K^{abmn} + K_{(S)}^{abmn} \,,
\label{3}
\end{equation}
$$
 K_{(S)}^{abmn} =
 \bar{\psi}\left[ Q_1 \gamma^a \gamma^b \gamma^m \gamma^n + Q_2 \gamma^a \gamma^m \gamma^b \gamma^n + Q_3 \gamma^a \gamma^n \gamma^b \gamma^m + Q_4 U^a U^b \gamma^m \gamma^n \right] \psi  +
 $$
 \begin{equation}
 + \frac{i}{2} Q_5 \left[\bar{\psi}\gamma^l \gamma^s \epsilon^{\ \ \ ab}_{ls}\gamma^m \gamma^n \gamma^5 \psi \right] + \frac{i}{2} Q_6 \left[\bar{\psi}\gamma^a \gamma^b \gamma^l \gamma^s \epsilon^{\ \ \ mn}_{ls}\gamma^5 \psi \right] \,.
\label{4}
\end{equation}
Here $\epsilon^{jkmn}=\frac{E^{jkmn}}{\sqrt{-g}}$ is the Levi-Civita pseudotensor with $E^{0123}=+1$.
Formally speaking, the new terms in (\ref{4}), with $Q_1$, $Q_2$, $Q_3$ and $Q_4$ in front, are the spinor analogs of the terms in (\ref{2}) with $C_1$, $C_2$, $C_3$ and $C_4$, respectively. The terms with $Q_5$ and $Q_6$ have no analogs in the canonic dynamic aether theory; in fact, these two terms involve into consideration the anti-symmetric dual tensor $\left(\nabla_m U_n \right)^* = \frac12 \epsilon_{mn}^{\ \ \ ls} \nabla_l U_s$.

The matrices $\gamma^j$ in the Riemann space-time with the metric $g_{pq}$ are connected with the constant Dirac matrices $\gamma^{(a)}$ defined in the Minkowski spacetime with the metric $\eta_{(a)(b)}$ via the tetrad vectors $X^j_{(a)}$, $\gamma^k = X^k_{(a)} \gamma^{(a)}$. The following relationships are used in this context. First, we deal with the normalization of the tetrad four-vectors:
\begin{equation}
g_{mn}X^{m}_{(a)}X^{n}_{(b)}=\eta_{(a)(b)} \,, \quad \eta^{(a)(b)}X^{m}_{(a)}X^{n}_{(b)}=g^{mn} \,.
\label{X}
\end{equation}
Second, the Dirac matrices satisfy the fundamental anti-commutation relations
\begin{equation}
\gamma^{(a)} \gamma^{(b)} {+} \gamma^{(b)}\gamma^{(a)} =2 E \eta^{(a)(b)}  \Leftarrow \Rightarrow \gamma^m \gamma^n   {+}  \gamma^n \gamma^m =2 E g^{mn} \,,
\label{comm1}
\end{equation}
where $E$ is the four-dimensional unit matrix.
Third, the link between the Dirac matrices $\gamma^5$ and $\gamma^{(5)}$ is
$$
\gamma^5 = -\frac{1}{4!} \epsilon_{mnpq} \gamma^m \gamma^n \gamma^p \gamma^q  =  -\frac{1}{4!} \epsilon_{mnpq} X^m_{(a)} X^n_{(b)} X^p_{(c)}X^q_{(d)} \gamma^{(a)} \gamma^{(b)} \gamma^{(c)} \gamma^{(d)} =
$$
\begin{equation}
{=} {-}\frac{1}{4!} \epsilon_{(a)(b)(c)(d)} \gamma^{(a)} \gamma^{(b)} \gamma^{(c)} \gamma^{(d)}  {=}
\gamma^{(0)} \gamma^{(1)}\gamma^{(2)}\gamma^{(3)} \equiv \gamma^{(5)}.
\label{gamma85}
\end{equation}
Fourth, the Fock - Ivanenko connection coefficients are used in the form
\begin{equation}
\Gamma_{k}=\frac{1}{4}g_{mn}X^{(a)}_{s}\gamma^{s}\gamma^{n} \nabla_{k}X^{m}_{(a)} \,.
\label{5D}
\end{equation}

\section{Master equations of the extended theory}

\subsection{Master equations of the unit vector field}

Variation of the action functional (\ref{1}) with respect to aether velocity $U^j$ gives the standard balance equations \cite{J1}
\begin{equation}
\nabla_a {\cal J}^{aj} = \lambda U^j + {\cal I}^j \,,
\label{j1}
\end{equation}
where the Jacobson tensor has now the extended form
\begin{equation}
{\cal J}^{aj} = \left[K^{abjn} + K_{(S)}^{abjn} \right]\nabla_b U_n \,,
\label{j2}
\end{equation}
and the Jacobson current contains new spinor term
\begin{equation}
{\cal I}^j = C_4 DU_m \nabla^j U^m + Q_4 DU_m \nabla^j U_n \left(\bar{\psi} \gamma^m \gamma^n \psi \right) \,.
\label{j3}
\end{equation}
Here $D=U^k \nabla_k$ is the convective derivative, and the Lagrange multiplier $\lambda$ can be standardly presented as follows
\begin{equation}
\lambda = U_j \left(\nabla_a {\cal J}^{aj} - {\cal I}^j \right) \,.
\label{j4}
\end{equation}

\subsection{Master equations of the spinor  field}

Variation of the action functional (\ref{1}) with respect to $\bar{\psi}$ and $\psi$ gives two equations for the spinor field in the standard Dirac's form, respectively,
\begin{equation}
i \gamma^k D_k \psi = {\cal M} \psi \,, \quad i D_k \bar{\psi} \gamma^k = - {\cal M} \bar{\psi} \,,
\label{D1}
\end{equation}
where the matrix of the effective mass has now the following extended form
\begin{equation}
{\cal M} = m E + \frac{1}{2\kappa} \nabla_a U_m  \nabla_b U_n \left\{\left[ Q_1 \gamma^a \gamma^b \gamma^m \gamma^n + Q_2 \gamma^a \gamma^m \gamma^b \gamma^n + Q_3 \gamma^a \gamma^n \gamma^b \gamma^m + Q_4 U^a U^b \gamma^m \gamma^n \right] + \right.
\label{M}
\end{equation}
$$
\left.
+ \frac{i}{2} Q_5 \left[\bar{\psi}\gamma^l \gamma^s \epsilon^{\ \ \ ab}_{ls}\gamma^m \gamma^n \gamma^5 \psi \right] + \frac{i}{2} Q_6 \left[\bar{\psi}\gamma^a \gamma^b \gamma^l \gamma^s \epsilon^{\ \ \ mn}_{ls}\gamma^5 \psi \right] \right\} \,.
$$
We see that the extension of the kinetic part of aether field modifies essentially the effective mass of the spinor particle.

\subsection{Master equations of the gravity field }

Variation  of the action functional with respect to metric $g^{pq}$ gives the Einstein equations in the standard form
\begin{equation}
R_{pq} - \frac12 g_{pq} R = \Lambda g_{pq} + \kappa T^{(D)}_{pq}  + T^{(A)}_{pq}  \,.
\label{G1}
\end{equation}
The first term in the right-hand side of (\ref{G1}) is
the canonic part of the stress-energy tensor of the spinor field
\begin{equation}
T^{(\rm D)}_{pq} = {-} g_{pq} \left[\frac{i}{2}(\bar{\psi}\gamma^{k}D_{k}\psi {-}D_{k}\bar{\psi}\gamma^{k}) {-} m \bar{\psi}\psi \right]  {+}
 \frac{i}{4}\left[\bar\psi \gamma_{p} D_{q}\psi {+} \bar\psi \gamma_{q} D_{p}\psi {-} (D_{p}\bar\psi) \gamma_{q} \psi {-} (D_{p}\bar\psi) \gamma_{q} \psi \right] \,.
\label{G3}
\end{equation}
The second term describes
the extended stress-energy tensor of the aether field interacting with the spinor field
\begin{equation}
T^{(A)}_{pq} =
\frac12 g_{pq} \ {\cal K}^{abmn} \nabla_a U_m \nabla_b U_n{+} U_p U_q  U_j \left(\nabla_a {\cal J}^{aj} - {\cal I}^j \right) {+}
\nabla^m \left[U_{(p}{\cal J}_{q)m} {-}
{\cal J}_{m(p}U_{q)} {-}
{\cal J}_{(pq)} U_m\right]+
\label{G2}
\end{equation}
$$
+C_1\left[(\nabla_m U_p)(\nabla^m U_q) {-}
(\nabla_p U_m )(\nabla_q U^m) \right] {+}
\left(C_4 + Q_4 \bar{\psi}\psi \right) DU_{p}DU_{q} -
$$
$$
-\frac12 \nabla_a U_m \nabla_b U_n \bar{\psi}\left\{ Q_1 \left[\gamma_{(p}\delta^a_{q)} \gamma^b \gamma^m \gamma^n + \gamma^a \gamma_{(p}\delta^b_{q)} \gamma^m \gamma^n  - \gamma^a \gamma^b \gamma_{(p}\delta^m_{q)}\gamma^n - \gamma^a \gamma^b \gamma^m \gamma_{(p}\delta^n_{q)} \right] + \right.
$$
$$
\left. + Q_2 \left[\gamma_{(p}\delta^a_{q)} \gamma^m \gamma^b \gamma^n - \gamma^a \gamma_{(p}\delta^m_{q)} \gamma^b \gamma^n  + \gamma^a \gamma^m \gamma_{(p}\delta^b_{q)}\gamma^n  - \gamma^a \gamma^m \gamma^b \gamma_{(p}\delta^n_{q)} \right]+ \right.
$$
$$
\left. + Q_3 \left[\gamma_{(p}\delta^a_{q)} \gamma^n \gamma^b \gamma^m - \gamma^a \gamma_{(p}\delta^n_{q)} \gamma^b \gamma^m  + \gamma^a \gamma^n \gamma_{(p}\delta^b_{q)}\gamma^m - \gamma^a \gamma^n \gamma^b \gamma_{(p}\delta^m_{q)} \right] \right\}\psi -
$$
$$
-\frac{i}{4} g_{pq}\nabla_a U_m \nabla_b U_n \bar{\psi} \left[Q_5 \epsilon^{lsab} \gamma_l \gamma_s \gamma^m \gamma^n  -
Q_6 \epsilon^{lsmn} \gamma^a \gamma^b \gamma_l \gamma_s  \right]\gamma^5\psi +
$$
$$
+ \frac{i}{4} \nabla_a U_m \nabla_b U_n \bar{\psi} \left\{Q_5 \epsilon_{ls}^{\ \ ab}\left[\gamma_{(p}\delta^l_{q)} \gamma^s \gamma^m \gamma^n + \gamma^l \gamma_{(p}\delta^s_{q)} \gamma^m \gamma^n + \gamma^l \gamma^s \gamma_{(p}\delta^m_{q)} \gamma^n + \gamma^l \gamma^s \gamma^m \gamma_{(p}\delta^n_{q)} \right] - \right.
$$
$$
\left. - Q_6 \epsilon_{ls}^{\ \ mn} \left[\gamma_{(p}\delta^a_{q)} \gamma^b \gamma^l \gamma^s + \gamma^a \gamma_{(p}\delta^b_{q)} \gamma^l \gamma^s + \gamma^a \gamma^b \gamma_{(p}\delta^l_{q)} \gamma^s + \gamma^a \gamma^b \gamma^l \gamma_{(p}\delta^s_{q)} \right] \right\} \gamma^5 \psi \,.
$$
We used here the following additional formulas
\begin{equation}
\delta \gamma^k = \frac12 \gamma_{(p}\delta^k_{q)} \delta g^{pq} \,, \quad \delta \gamma_k = -\frac12 \gamma_{(p}g_{q)k} \delta g^{pq} \,, \quad
\delta \gamma^5 =0 \,, \quad \delta \epsilon^{lsmn} = \frac12 \delta g^{pq} g_{pq} \epsilon^{lsmn} \,.
\label{G21}
\end{equation}
As usual, the parentheses denote the symmetrization $A_{(p}B_{q)} = \frac12 (A_pB_q + A_q B_p)$.

The procedure of the derivation of the total set of coupled master equations for the unit vector field, spinor and gravitational fields is completed.

\section{Outlook}

The extended model of interaction between the unit vector field, spinor and gravitational fields, which takes into account the backreaction of the spinor field on the dynamic aether,  is established. The total set of the corresponding coupled master equations is derived. What is the next step in the development of this theory? We expect that this theory will give interesting results in application to cosmology and astrophysics. These investigations are planned to be realized in the nearest future.


\end{document}